# Presentation a Trust Walker for rating prediction in Recommender System with Biased Random Walk: Effects of H-index Centrality, Similarity in Items and Friends

Saman Forouzandeh[1], Mehrdad Rostami[2], Kamal Berahmand[3]

Department of Computer Engineering University of Applied Science and Technology, Center of Tehran Municipality ICT org.Tehran, Iran[1]
Department of Computer Engineering, University of Kurdistan, Sanandaj, Iran[2]
Department of Science and Engineering, Queensland University of Technology, Brisbane, Australia[3]
Saman.forouzandeh@gmail.com[1], m.rostami@eng.uok.ac.ir[2], kamal.berahmand@hdr.qut.edu.au[3]

## Abstract

The use of recommender systems has increased dramatically to assist online social network users in the decision-making process and selecting appropriate items. On the other hand, due to many different items, users cannot score a wide range of them, and usually, there is a scattering problem for the matrix created for users. To solve the problem, the trust-based recommender systems are applied to predict the score of the desired item for the user. Various criteria have been considered to define trust, and the degree of trust between users is usually calculated based on these criteria. In this regard, it is impossible to obtain the degree of trust for all users because of the large number of them in social networks. Also, for this problem, researchers use different modes of the Random Walk algorithm to randomly visit some users, study their behavior, and gain the degree of trust between them. In the present study, a trust-based recommender system is presented that predicts the score of items that the target user has not rated, and if the item is not found, it offers the user the items dependent on that item that are also part of the user's interests. In a trusted network, by weighting the edges between the nodes, the degree of trust is determined, and a TrustWalker is developed, which uses the Biased Random Walk (BRW) algorithm to move between the nodes. The weight of the edges is effective in the selection of random steps. The implementation and evaluation of the present research method have been carried out on three datasets named Epinions, Flixster, and FilmTrust; the results reveal the high efficiency of the proposed method.

*Keywords:* Recommender Systems; Biased Random Walk; TrustWalker; Association Rules

## 1. Introduction

The expansion and investment of various commercial and service companies in social networks have caused users to encounter a new challenge called information overload, which means that a great deal of information and items have been provided to users, and this is the reason that they may not be able to make choices in accordance to their needs and interests [1, 2]. In order to solve this challenge, the use of recommender systems has become prevalent in recent years. Recommender systems are employed to recommend a variety of items to users by applying statistical techniques and knowledge discovery [3]. Recommender systems attempt to make suggestions to users according to their performance, personal tastes, user behaviors, and the context in which they are applied to match their personal preferences and assist them in the decision-making process [4-6]. In order to provide recommendations to users, usually, a matrix is formed that contains some items that the user has ranked; however, the problem is related to the high dispersion of these matrices, and due to a large number of available items, the user cannot rank a large number of them and must predict their rankings [7, 8].

In the real world, users are often influenced by people they trust, and therefore the recommendation made by trusted people is far more influential than the suggestions of other people. In these systems, the lack of data related to item ranking is compensated by social relations [9]. Accordingly, one of the challenges is that users usually do not rank a large number of items and predict the ranking of these specific items and finally provide related recommendations. To predict the ranking of items, in recent years, researchers have used trust-based recommender systems and have proposed various approaches to calculate trust between users and make different recommendations based on this trust. In general, trust is of two types; it can be gained directly by the users (Explicit Trust) or obtained from their behavior (Implicit Trust)[10]. Since explicit statements are not always available, implicit techniques are more practical compared to explicit approaches [11]; hence, implicit trust is applied in the present investigation. The development of a trust relationship between users can possess different reasons, and usually, in social networks, the user's neighbors can be considered as excellent options for users trusted by the target user [12]. Trust can be measured and assessed by a variety of criteria. Each of the works applies several criteria to assess the trust level between users, and the accuracy of the recommendations to users based on the trust level between them is consequently of great importance. In most cases, the calculation of trust was based on the similarity of users in selecting items or their neighbors. The use of this type of trust may not be significantly accurate in making recommendations since only behavioral similarity, or neighborship cannot be a reliable factor in creating trust and the user that is considered as a trusted user may not be highly ranked in the social network (have few friends) and may not be a known and famous person, and this issue causes distrust. In the present research, in order to solve this challenge, unity in the number of friends, as well as the importance of the user in the social network, is applied in addition to behavioral similarity in selecting similar items and in order to measure the importance of the user, a centrality is defined for the user by applying H-index, and this index and its improvement are used for the case that the node degrees have not reached the desired threshold to evaluate the importance of the user in forming a trust relationship between them and the target user. Another controversial issue is the numerous users in Online Social Networks (OSN) that sometimes it is practically impossible to evaluate thousands of users (user's friends and also their friends, etc.) to calculate the trust level of the target user and most researchers apply Random Walk method to solve this problem [13, 14].

A number of users are selected, and the target user's trust in them is calculated and evaluated using Random Walk and creating random steps. The Random Walk output may also provide a great deal of irrelevant data, and the movement of random steps in-depth network should also be limited. In order to solve these problems, researchers have primarily developed a trusted network for the source user, which includes users who possess basic criteria (e.g., unity in the selection of items, neighbors, etc.) to trust the target user in them [15, 16]. The random Walk length is also limited [15]. However, Random Walk also possesses uniformly (or almost-uniformly) selections, and it is needed to do something to increase the probability of selecting the required users and reduce the possibility of selecting distinct users with defined criteria. The Biased Random Walk (BRW) is used in the present study; the probability of selecting users can be increased by weighting the edges between nodes that are more similar to the desired criteria.

In the present research, the authors primarily develop a trusted network based on the desired criteria for the target user, then examine the target user's behavior and seek items in the trusted network that the user has not ranked. In order to find the ranking of the desired item in the trusted network, a Trust Walker is used, and the search is made using Biased Random Walk (BRW), and the movement of Trust Walker is based on random steps from the source user, with the difference that contrary to the standard Random Walk in BRW, the weight of the edges affects the selection of random steps. The weight of the edges between users determines how much they trust each other in the trusted network, and the higher this weight, the trust level increases. The way of measuring trust between users has also been based on three criteria: similarity in the selection of items, the similarity in connections or mutual friends between users, and the centrality of the H-index (to determine the prominence of users in social networks). The proposed recommender system functions in such a way that if the visited user has rated the item, it returns the ranking of that item to the user, and if it has not ranked the item, it suggests items that are most dependent on the target user's favorite items and have the most repetition by evaluating the behavioral share of users visited by the source user as well as applying the association rules.

The rest of the present paper is organized as follows. The second section discusses related work. The third section contains the problem definition, and the fourth section addresses the proposed method. In the fifth section, the method is evaluated, and in the sixth section, the conclusion is made.

## 2. Related Work

In recent years, due to the increasing volume of data and the variety of services provided to users, the selection of suitable items for users has become challenging, and a variety of recommender systems have been developed. The applied types of systems include Content-based Filtering, Collaborative Filtering, or a combination of them. To provide personalized recommendations, there are two ways to capture users' preferences [17]: implicit and explicit. In implicit feedback [18], the system infers userâs preferences by monitoring different actions of users such as purchasing history, browsing history, clicks, email contents, etc., so this type of feedback reduces the burden from the user. In explicit feedback [19], recommender systems prompt users to provide ratings for items to reconstruct and improve their model. In the section of related works, first, a review of Collaborative Filtering recommender systems is performed, then trust-based recommender systems are evaluated, and finally, several types of approaches related to Random walk and its application in trust-based recommender systems are analyzed.

## 2.1. Collaborative Filtering

The use of Collaborative Filtering methods has proven effective because of using the participating rate of users in making recommendations, especially in the case of applying hybrid methods. The collaborative Filtering recommender systems employ user preferences to rank the items, which means that users detect a similar target user, and ranking them by these users is considered an objective for the user [14].

Collaborative filtering methods are further divided into three categories: memory-based, model-based, and hybrid of both [20-22]. Memory-based methods utilize users' past behavior and recommend products that other users with similar interests have selected in the past [23, 24]. They have been widely used in commercial recommender systems [25]. Memory-based algorithms are either user-based [26] or item-based [27, 28]. User-based algorithms predict rating given by a user to an item based on the ratings by similar users, whereas, item-based algorithms estimate the rating based on the ratings of similar items previously chosen by the user. Methods used in traditional recommender systems are mostly based on user-item rating matrix. Model-based methods utilize available data to train a predefined model for rating prediction. Some of the commonly used: clustering [29] and Matrix Factorization model [30]. Model-based approaches can handle problems with limited data using hierarchical clustering to enhance the accuracy of the prediction [30, 31]. Matrix factorization factorizes the user-item rating matrix using low-rank representation. Although model-based methods mitigate the sparsity problem, handling users who have never rated any item is a challenging problem in both memory-based and model-based approaches.

## 2.2. Trust Recommender Systems

Simultaneously with the growth of the application of various social networks, the use of recommender systems has been equally considered by researchers. Over the past few years, the relationship between users in social networks has been studied from different aspects, and one of these perspectives is the relationship between users based on the trust level between them. Many studies have indicated that the selection and purchase of different items have been made based on the recommendations of friends and people who were trustful for the user [16]. Accordingly, if the interactions between users on social networks are based on trust between them, the efficiency of recommender systems in providing a variety of suggestions to users is increased, and it will be promising that the target user to receive some suggestions from more trusted users; thus, the possibility of accepting the recommendations increases [32]. The more trust between the two users, the higher the percentage of acceptance of the target user's recommendations. In this regard, many pieces of research have been conducted related to trust recommender systems [8, 33, 34].

Trust methods can be classified into implicit and explicit methods. Implicit trust is usually obtained from user-item interactions (i.e., ratings), and explicit trust is extracted from the user relationships (who they trust and up to what extent) [20, 35-37]. These methods analyze pre-existing relationships in a web of trust for an active user [38, 39]. Collaborative filtering methods are most effective when users have expressed enough ratings.

The use of these types of trust usually predicts the ranking or suggestion of items to the user and is often applied in cases where the considered items possess a ranking of 1-5. In [40], a method based on trust and matrix factorization has been presented called TrustSVD, which uses explicit and implicit types of trust to predict not-ranked items by users and is unknown for them. In [41], a recommender system based on Mobile Applications is presented, which calculates the similarity between users according to social relationships and the trust level between them; this method is used to solve the data sparsity and cold start problems. In order to establish trust, users' social relationships and interactions, as well as their social reputation on the social network, are used.

## 2.3. Trust-based Random Walk

The basis of all recommender systems is identical. Their similarity is in assigning and recommending items of interest to users. The use of random walking in recommender systems has increased significantly in recent years and was first mentioned in [42] and also used in the book recommender system in [43]. In [44] a random walk has also been applied to recommend a movie on YouTube. The TrustWalker algorithm was first proposed by Jamali [15], which was a combination of a trust-based method and a commodity-based group refinement method. In this method, a random walk has been used to refine the trusted network, and it has been indicated that people who are closer to the user and have ranked a similar user-item are more valuable compared to users who ranked the user-item but are farther away from the user (they are not the user's direct neighbor).

There are some algorithms such as Eigentrust [45], Appleseed [46], and another algorithm in [47] that use principal eigenvector to make trust computations. However, these methods produce ranks of trustworthiness of users, so they would be suitable for systems where ranks are considered. The TidalTrust model finds all raters with the shortest path from the source user and aggregates their ratings weighted by the trust between them. Another method is MoleTrust [48] where the computation of trust value between two users is based on backward exploration. Also, trust values in recommender systems help to predict the behavior of those users who have rated fewer products [49]. In the paper [50], authors propose a hybrid recommendation technique that combines content-based, collaborative filtering, and random walk with restart method recommendation algorithm by utilizing social network information. In [17], a method called ContextWalk has been proposed in which the movie recommender system on the related website modeled the user's behavior on the contextual graph according to the user's selections and by applying random walking. In [18], an algorithm based on random walking was developed in which the problem of cold start is solved for users in social tagging data.

The potential transitive relations between users and items can be captured through their interaction with tags. In [19], the authors provide a random walking ranking algorithm called ItemRank, which is capable of ranking users' preferences and providing high-ranking items and recommendations. Authors in [51] have presented a method based on a random walk with LDA. They use LDA to mining the potential property of products and apply the results to construct the transition matrix for the random walk. They analyzed the structure of the networks through the distribution of the degree of the nodes and discussed how they influence the performance of the algorithms.

## 3. Problem Definition

In recommender systems, there is a set of users in the form of $U = \{U_1, U_2, ..... U_N\}$ and they rate some items in the form of $I = \{I_1, I_2, ..... I_M\}$; thus, we have: $RIu = \{I_{u1}, I_{u2}, ...., I_{uk}\}$. The score of item I by user U is in the form of ru,i, which is often a number in the range of 1 to 5; the larger the number, the more the user is interested in that particular item. In trust-based systems, the trust or distrust between users is often indicated by 0 or 1. If node U trusts node V, $t_{u,v}$ equals 1, and a value of 0 indicates the distrust of node $U$ to node $V$; this type of trust is known as binary trust, e.g., Amazon and eBay. In this regard, if $TU_u = \{v \, \mathcal{E} \, U \, | t_{u,v} = 1\}$ that represents the set of users that are directly trusted by u, the trusted network containing a graph can be defined as: $G = <U, TU>$ where $TU = \{(u, v)|u \, \mathcal{E} \, U; v \, \mathcal{E} \, TUu\}$. There is one node for each user, and the number placed between the nodes on each edge indicates the degree of trust of the two nodes with a friendly relationship in the trusted network. In general, a user ($u \, \mathcal{E} \, U$) and an item ($i \, \mathcal{E} \, I$) are given and $r_{u,i}$ is unknown for a recommender system, which means that the recommender system must be capable of predicting the score of the item $I$ rated by user $U$. The score predicted by the recommender system is indicated by $r'_{u,i}$. The conventional recommender systems usually obtain the prediction of the score of item ŕu, i using the similarities between users [33]. They find users who have rated the desired item for the target user, select users among them who have similar profiles to the target user and predict the score of the considered item in this way. In trust-based recommender systems, item score prediction is performed using behavior analysis and profiles of trusted users of the target user by developing a trusted network; this means that it checks which of the trusted users of the target user has rated that item to assign a score to the item. Because of the large number of users in online social networks, it is impossible to check all users and calculate the degree of trust of each user. To solve this problem, most researchers use the Random Walk algorithm and randomly select some users and calculate the degree of trust based on the desired criteria among users.

In the present study, in addition to trust-based relationships between users based on the degree of similarity between them in selecting items, their common connections (the number of mutual friends), and the importance of users in social networks are used to predict the score of items. Therefore, a pattern of social trust is developed between users that both the importance of users in social networks and the degree of similarity between them (based on common items and connections) are considered. In the following, a TrustWalker is created using Biased Random Walk (BRW) that starts to move from the source node and is randomly affected by the weight placed on the edges (according to the criteria) in the selection of nodes, and then checks whether the selected user has rated the source or target user's considered item. If the visited user has rated the item, its score will be returned; otherwise, Trust Walker selects a node based on the weight of the edges and checks whether the visited user has rated that item or not by affecting by them, and based on the average distance achieved for each dataset, the level of TrustWalker movement in the trusted network is obtained, and if the score of the desired item is not found among the visited user, the behavior of these users with the target user is assessed using association rules and some items are recommended to the target user that are related to the items rated by them as well as their scores are in the range of scores related to the items of interest to the target user. If the score of searched items is not performed by visited, the items related to their interests are offered to the target user; hence, the similarity criterion between users in selecting similar items is more important than the other two criteria. Accordingly, the impact factor of item similarity in weighting the edges is considered to be double; therefore, the probability of selecting users who have item similarity with the target user is increased by BRW.

Two matrices are used to develop a trusted network. In the first matrix, users are searched who have rated and selected common items with the target user, and the second matrix includes the users who have mutual friends with the target user. Then, the desired graph or trust network is created between the users by combining these two matrices. In the following sections, first, the development of the trusted network is discussed.

### 3.1. Trust Network Construction
In this section, a trusted network is developed for the target user. The purpose of creating a trusted network is to find users who are trusted by the target user according to the defined criteria, and these users are identified in two steps and then combined, and the relationship between them is determined, and finally, a unit graph is created, which is considered as a trusted network.

### 3.1.1. Creating a Matrix Based on the Similarity of Items
To discover users who have rated items similar to the target user, a user-item matrix is created, and users are placed in each row of the matrix, and the matrix columns represent the rated items by users. An example of a user-item matrix is shown in Table 1, in which the users (U) rate their desired items (I) and the similarity of users with the target user in the selection of desired items is determined through this matrix, and each user with behavioral similarity with the target user enters the considered trust graph. The user-item matrix is presented in Table 1.

Table 1. The user-item Matrix

|       | $I_1$ | $I_2$ | $I_3$ | $I_4$ | $I_5$ | ... |
|-------|-------|-------|-------|-------|-------|-----|
| $U_1$ | 3 | 5 | - | 4 | 2 | ... |
| $U_2$ | 3 | 4 | 2 | - | 4 | ... |
| $U_3$ | - | 1 | - | 3 | 3 | ... |
| $U_4$ | 2 | 5 | 3 | 4 | 5 | ... |
| $U_5$ .... | - | - | 2 | - | 5 | ... |

### 3.1.2. Creating a Matrix Based on Friend Similarity (Sharing Links and Connections)
In order to detect users with mutual friends with the target user, a user-user matrix is created, and the desired users are placed in each row of the matrix, and the matrix columns represent the friends of the considered user. Accordingly, if each user has a friendship with another user, the number 1 will be written in the desired matrix entry; otherwise, 0 will be placed between two users. For instance, in Table 2, user *U1* has a friendly relationship with users *V2*, *V3*, and *V4*, and the number 1 is placed in front of each of the columns attributed to these users. Similarly, through this matrix, each user's relationship with other people is determined. The user-user matrix is presented in Table 2.

Table 2. The user-user matrix

|       | $V_1$ | $V_2$ | $V_3$ | $V_4$ | $V_5$ | ... |
|-------|-------|-------|-------|-------|-------|-----|
| $U_1$ | 0 | 1 | 1 | 1 | 0 | ... |
| $U_2$ | 1 | 0 | 1 | 1 | 1 | ... |
| $U_3$ | 1 | 1 | 0 | 0 | 1 | ... |
| $U_4$ | 1 | 1 | 0 | 0 | 1 | ... |
| $U_5$ .... | 0 | 1 | 1 | 1 | 0 | ... |

In the following, according to the outputs of Tables 1 and Table 2 for each user, the users who have similarity with them in the selection of items or have mutual friends with the user are identified; finally, they are placed in a graph named "trust network" and their relationships between with the target user and other users is specified.

## 3.2. Definition Trust

After creating a trusted network, some criteria are introduced that the target user's trust in other users in the trusted network is based on these criteria. The criteria considered to define trust in the present study include calculating two types of similarities between users and defining a centrality for each user to assess the importance of the user in the social network. The criteria considered to define the desired trust between users are as follows.

### 3.2.1. Similarity-based Trust
The similarity between users is calculated based on two factors as the following.

### 3.2.1.1 Rating Similarity
In this section, the similarity of users in selecting common items is calculated. To calculate the similarity between users in selecting the items rated by both users a and b, the Pearson correlation coefficient function is used as follows:

$$Sim_{Pea}(a,b) = \frac{\sum_{i \in A_{a,b}} (r_i(a) - r_i^{'}(a)) \cdot (r_i(b) - r_i^{'}(b))}{\sqrt{\sum_{i \in A_{a,b}} (r_i(a) - r_i^{'}(a))^2} \cdot \sqrt{\sum_{i \in A_{a,b}} (r_i(b) - r_i^{'}(b))^2}} \cdot \quad (1)$$

Accordingly, $r_i(b)$ and $r_i(a)$ represents the scores respectively given by the users a and b for the item i. Also, $r^{'}(b)$ and $r^{'}(a)$ represents the average scores given by the users a and b, respectively. $A_{a,b}$ represents a set of items rated by both users a and b. The values returned by the Pearson function are in the range of [-1,1]. The negative correlation indicates that there is no similarity between the two users in the selection of items. Only the positive values returned by the Pearson function are considered in the present investigation, which indicates the similarity between users in the selection of items. The way of calculating the impact factor for this level of similarity between two nodes is that the sum of the items rated by the two users is calculated and the result is divided on the number of common items between two users; thus, the following relation is obtained:

$$Sim_{Item}(a,b) = \frac{\sum_{i \in A_{a,b}} (r_i(a) + r_i^{'}(b))}{r_i(a) \cap r_i^{'}(b)} \quad (2)$$

As can be seen, the sum of the items rated by the two users a and b and the obtained result is divided into the common rated items between the two users, and the outcome is considered as the impact factor of the similarity of items between two users.

### 3.2.1.2. Connection (Friends) Similarity
In this section, users' similarity is calculated based on the number of similar connections or mutual friends. In this regard, in addition to measuring the similarity of items, another function is introduced to calculate the similarity of users in their connections, by which the users' links or mutual friends can be detected. Thus, the following equation can be written:

$$Sim_{Con}(a,b) = \frac{F(a) \cap F(b)}{F(a)} \quad (3)$$

Where, the number of mutual friends between the users a and b is calculated in relation to the total number of friends of the user a, and the larger this number is, the two users have higher similarity in their friends and connections. The considered impact factor is also measured based on the following equation:

$$Sim_{Deg}(a,b) = \frac{(\sum a_{ij} = a_{ji}) + (\sum b_{ij} = b_{ji})}{(\sum a_{ij} = a_{ji}) \cap (\sum b_{ij} = b_{ji})} \quad (4)$$

This formula calculates the sum of the degrees of two nodes and divides the result between the links or mutual friends between two nodes; the final result is equal to the considered impact factor.

### 3.3. Centrality-based Trust

The users with high importance (high impact) represent the degree of user's influence on other users [52], and various criteria have been defined to calculate this level of influence; in the present research, the H-index Centrality criterion is used for this purpose. In a general definition of H-index centrality, a threshold (k) is considered, which includes neighbors with minimum degrees of k for each node. Accordingly, the function $C_k(V_i)$ was first defined, including several neighbors of the node $V_i$ with the degrees greater than or equal to k. Hence, the following equation is obtained:

$$c_k(v_i) = |\{v_j | v_j \in N_i \text{ and } d_j \geq k\}| \quad (5)$$

In the following, the H-index function can be defined as follows [53]:

$$H - \text{index}(v_i) = max_k(c_k(v_i)) \quad \text{where} \quad c_k(v_i) \geq k \quad (6)$$

The function's value is a maximal value for k such that k of the neighbors has degrees greater than or equal to k. The considered value for k is equal to the average degree of neighbors of the node $v_i$, and for example, if the average degree of neighbors of the node $v_i$ is equal to 3, only the nodes get impact factor that the minimum degree of which equals to higher than or equal to 3. The average degree of neighbors of one node is also obtained according to Eq. (7).

$$\text{Ave\_n\_Deg}(i) = \left\lfloor \frac{\sum \deg(neigh \ v_i)}{\deg(v_i)} \right\rfloor \quad (7)$$

According to this equation, the average degree of neighbors is equal to the absolute floor of the total degree of neighbors of the node $v_i$ divided on the degree of $v_i$ where the obtained number is considered as the threshold for K in the H-index formula. The way of applying H-index is that an impact factor is added for each neighbor with a degree equal or higher than the average degree of the considered neighbor of the node, and the value of the impact factor depends on the number of neighbors of each node. For instance, if the visited node of met $v_i$ possesses a degree of 10 and the average degree of its neighbors is equal to (k = 4), and three of the node's neighbors possess at least a degree of 4, then for each node with a minimum degree of 4, the value of 0.1 (1/10=0.1) is added to the node $v_i$i as the impact factor; therefore, the total impact factor of the visited node $v_i$ will be equal to (3×0.1=0.3). It is noteworthy that the H-index only selects nodes with a minimum degree of k; hence, some nodes with a degree less than the defined threshold (k) may be deleted, and a part of the structural information of the network can be ignored accordingly [54], while these nodes may be of great importance in the network.

In order to solve this problem in the present article, the H-index is enhanced, and the improved version is used. For example, consider the node $v_i$ that one of its neighbors is the node $v_j$, and the degree of node $v_j$ is lower than the obtained average for the degrees of neighbors of the node $v_i$. Therefore, the first-degree neighbors (direct neighbors) of the node $v_j$ are evaluated, and if the degree of each of the neighbors of the node $v_j$ is equal to the minimum average obtained for the degree of neighbors of the node $v_i$, an impact factor equal to $\frac{1}{2}$ is considered for the node $v_j$ for each neighbor of the node $v_j$, and accordingly, the value of the impact factor for each neighbor of the node $v_j$ with a minimum degree equal to the average of neighbors of the node $v_i$ is considered ($\frac{1}{2} \times 0.1 = 0.05$), that means that if the degree of node $v_j$ is equal to 3, but the degree of two of the neighbors of the node $v_j$ is higher than 3, the impact factor considered for node $v_j$ will be equal to (2*0.05 =0.1). Therefore, the following equations will be obtained as follows:

$$c_k(v_i) = |\{v_j | v_j \in N_i \text{ and } d_j < k \text{ and } v_j d_j < k \quad (8)$$

$$C_K(V_i) = |\{V_j | V_j \in N_i \text{ and } d_j < K \text{ and } V_j d_j \geq K\}| \quad (9)$$

Eq. (8) includes a node, the degree of which and its neighbors are less than k, and this node is considered as a nonsignificant node, and the impact factor is not considered as the node's importance by the H-index. Eq. (9)

includes the nodes with a degree less than k but the degree of node's neighbors is equal or higher than the value obtained for k, and therefore the number of this node's neighbors with a degree equal or higher than k is considered to be the impact factor with the value of $\frac{1}{2}$.

### 3.4. TrustWalker Model

In a graph, the relationship between nodes is usually indicated by the weight of edges between them. The idea of Random Walk is that the user starts moving from a specific node in the graph selects one of them with an equal probability in each step according to the number of neighbors of each node, and this action is repeated to reach the desired node. If G = (V, E) is a connected graph with n vertices and m edges. A Random Walk on the graph G is as follows: starting from the node $v_0$, if we are in the $t_{th}$ step of the node $v_t$, we move toward one of the neighbors of $v_t$ ($d(v_t)$ is the degree of the node $v_t$) with the probability of $\frac{1}{d(v_t)}$. The sequence of random nodes of $v_t$ t = 0, 1, 2,… is a Markov chain. The node $v_t$ can be assumed to be constant or selected by the distribution of $P_0$ among the graph nodes. The distribution of $v_t$ is denoted by $P_t$ as follows:

$$P_t(i) = \text{Prob } (V_t = i) \quad (10)$$

The transfer probability matrix in the Markov chain is indicated $M = (p_{i,j})_{i,j \in V}$ as the following:

$$(p_{i,j}) = \begin{cases} \frac{1}{d(i)} & \text{if } ij \in E \\ 0 & O \cdot W \end{cases} \quad (11)$$

Now it is supposed that $A_G$ is the adjacency matrix of the graph G, and also D is the diagonal matrix ($D_{i,j} = 0, i \neq j$), which is defined as follows:

$$D_{i,j} = \frac{1}{d(i)} \quad (12)$$

Hence, the matrix M can be calculated with a simple matrix operator.

$$M = D \times A_G \quad (13)$$

This equation is applied when the purpose is to perform the Random Walk uniformly. However, if each of the edges possesses a weight (in the present study, each of the edges each has a weight), $A_G$ is the weighted matrix of the graph G and D is the diagonal matrix; hence, the following equation is obtained:

$$D_{i,j} = \frac{1}{\sum_{(i,j) \in E} W_{i,j}} \quad (14)$$

Then, using Eq. (10), the transfer probability matrix can be obtained for the state that the edges are weighted. The weight of edges between nodes indicates the probability of the selection of which node by the user. Random steps are modeled using the probability of transferring between nodes with different information such as user's preferences or related tags [17]. Generally, the traditional Random Walk algorithm has high precision but low personality and diversity [55].

In the present paper, the Random Walking algorithm is improved, and Biased Random Walk (BRW) is applied in which the movement and selection of random steps will be affected by the weight between the edges and each node with higher edge weight based on the considered criteria is more probably to be selected by TrustWalker. In network science, a biased Random Walk on a graph is a time path process in which an evolving variable jumps from its current mode to one of the various potential new modes, unlike in a pure Random Walk, the probabilities of the potential new modes are unequal [15].

The TrustWalker model proposed with the BRW algorithm is known as Centrality Connection Items-TrustWalker (CCI-TrustWalker). Each node is rated based on three defined criteria, and the higher the node's score, the more probable to be selected in the Random Walk. Distancing from the active user is one of the main challenges in the use of trust networks in recommender systems. In the trusted network, the more the distance from the active user, the wider the scope of information; however, the precision and importance of the information decrease. To solve this problem, the movement of TrustWalker in the trusted network is limited, and in each of the three studied datasets, the average distance is calculated, and the number obtained for them is approximately equal to 6; therefore, the maximum length of TrustWalker steps for the movement in the depth of the network is equal to six nodes. In Ref. [15], the same value has been considered for TrustWalker.

TrustWalker launches the Random Walk by starting from the active user $U_0$ and is placed in step k of the Random Walk at node U, then checks whether it possesses the desired item score or not; if there is no score for the item, it goes to the next node, and this action is repeated until it finds the considered score or reaches the maximum network depth. If the user visited by BRW has rated the item i, the walk is stopped, and the score of item i is set as $r_{i,u}$, and returned as the Random Walk output. However, if the item is not rated by this visited user or node, two states will occur as follows:

1. The walk is not continued with the probability of $P_{i,u,k}$, and two modes of A and B will be discussed.

A: The Random Walk is stopped and returns a number as output (0 or 1). In this case, the following conditions must be simultaneously satisfied.

A-1- The visited users not to score the item searched by the target user or source.

A-2- The length of Random Walk steps is calculated based on the average distance for each dataset, and the result is approximately equal to 6, and accordingly, the maximum number of TrustWalker steps in the trusted network is equal to six steps ($l = 6$).

A-3- After evaluating the behavior of the visited users of X, it should be determined that there is no behavioral similarity between the visited users and the source user in the selection of items, and according to Eq. (1), there is a negative correlation between two users, and the output of the numeric function is negative.

According to mode A, the number of k steps must be included in the calculation of $P_{i,u,k}$ so that long walks can be avoided, and if there are all three of the above conditions, TrustWalker returns a number as an output; so that it returns 1 for large inputs and 0 for small inputs as follows:

$$S = \frac{1}{1+e^{-t}} \quad (15)$$

According to Eq. (15):

$$P_{i,u,k} = W_{ij} \times \frac{1}{1+e^{-t}} \quad (16)$$

As indicated, the weight obtained based on the items as mentioned above is considered as the main criterion for stopping.

B: The mode B is also similar to mode A. Only in condition A-3, the output of the Pearson function of Eq. (1) is a positive number (positive correlation) and indicates the presence of common items between the visited users and the target user. In this case, the user's behaviors are evaluated by association rules, and the items dependent on the items rated by the target user are identified, and ultimately are recommended to the target user.

2. The walk is continued with the probability of $1 - P_{i,u,k}$, and the movement toward the next node is done, which is the neighbor of the current node U.

If the decision is to continue the walk, one of the neighbors of node U should be selected. If the variable $S_u$ is considered as a random variable leading to the random selection of the node V, which is the neighbor of node U, Eq. (17) will be obtained for $S_u$ as follows:

$$P(S_u = V) = \frac{W_{ij}}{\sum_{i \in TU_i} W_{ij}} \quad (17)$$

Where, $Wi$ indicates the weight between the nodes that is obtained using the desired criteria and includes Eq. (18).

$$\begin{cases} \alpha_1 = Sim_{Item}(a,b) \\ \alpha_2 = Sim_{Con}(a,b) \\ \alpha_3 = Cen_{H-index}(a,b) \end{cases} \Rightarrow W_{ij} = (2 \times \alpha_1) + \alpha_2 + \alpha_3 \qquad (18)$$

As mentioned earlier, the importance of item similarity ($\alpha_1$) is higher than the other two criteria and its impact factor is considered to be doubled because of the high probability of selecting users who have similar behavior with the target user so that if the searched item is not rated by users, another item could be recommended to them using the behavioral similarity between them and the target user. Accordingly, the proposed CCI-TrustWalker model is defined as Eq. (19):

$$CCI_{ij}^{BRW}(t) = \sum_{t=1}^{l} \frac{W_{ij}}{\sum_{j \epsilon \Gamma(i)} W_{ij}}(t) + \frac{W_{ji}}{\sum_{i \epsilon \Gamma(j)} W_{ji}}(t) \qquad (19)$$

Where ($l$) and $\Gamma(i)$ are denoted as the length of the path in the graph and the first-order neighborhood of a node, respectively. $W_{ij}$ contains the sum of the weights based on the parameters $\alpha_1, \alpha_2, \alpha_3$ (Eq. 18) and makes the CCI-TrustWalker movement biased according to the edge weights. The reason for the use of $\frac{W_{ij}}{\sum_{j \epsilon \Gamma(i)} W_{ij}}(t) + \frac{W_{ji}}{\sum_{i \epsilon \Gamma(j)} W_{ji}}(t)$ is that in the CCI-TrustWalker movement, given that the level of trust between nodes "i and j" and "j and i" may be different and the symmetry problem is solved accordingly, the levels of trust of "i and j" and "j and i" are summed.

### 3.5. Recommendation

The recommender system operates in two parts. In the first step, if the score of the considered item is in the trust network, it is detected and presented to the target user by the recommender system. The implementation of the CCI-TrustWalker compares the behavior of the visited node with the target user in terms of item similarity. If the item similarity is satisfied between two users, it goes to the next node, and it repeats this operation to depth 6 in the trusted network in the case that the desired node is not found. It then recommends items to the target user, which have not been rated by the target user. However, they rated the visited nodes, and these items are dependent on the common items between two users, and also the scores given to them by the visited users are in the range of target user's interests. Accordingly, first, the score that determines the target users' level of interest should be discovered; for this purpose, the average scores given by the user must be detected. Hence, the following equations can be stated:

$$\text{Interests} = X \geq \sum_{i \epsilon n}^{i=n} \left\lfloor \frac{r_{ui}}{n} \right\rfloor \qquad (20)$$

$$\text{Lack of Interest} = X < \text{Interests} \qquad (21)$$

In Eq. (20), $r_{ui}$ (item u) is rated by the user r and given a score i by the user. n is referred to as the number of rated items. The value of i is equal to the number of items rated by the user. Since the scores are integers, the average obtained from the user's rating is placed within the absolute floor. Accordingly, if the score given to an item suggested by the visited user is equal or higher than the average user scores, it indicates the user's interest in that item, and if it is less than the value obtained for interests, it shows the user's lack of interest in that particular item. The association rules are applied to assess the user's behavior and identify repetitive and interdependent items. The used association rules consist of three items, as follows [5, 56]:

1. Support: The ratio of the number of intersections in which both objects A and B are present, to the total number of records. The value obtained for support is in the range of 0 to 1, and the greater values indicate the higher relationship between these two objects. The support criterion for A and B is shown in Eq. (22) and Eq. (23) as follows:

$$Supp\ (A \Rightarrow 0) = P\ (A \cap 0) = P\ (A)\ P\ (0\ |\ A) = P\ (0)\ P\ (A\ |\ 0) \qquad (22)$$

$$Supp\ (B \Rightarrow 1) = P\ (B \cap 1) = P\ (B)\ P\ (1\ |\ B) = P\ (1)\ P\ (B\ |\ 1) \qquad (23)$$

2. Confidence: The value of this criterion is also in the range of 0 to 1, and higher values will increase the quality of added rule. The use of this criterion, along with support, would be an excellent complement to evaluate the association rules. The confidence criterion for A and B is presented in Eq. (24) and Eq. (25) as follows:

$$\text{Conf}(A \Rightarrow 0) = P(A|0) \quad (24)$$

$$\text{Conf}(B \Rightarrow 1) = P(B|1) \quad (25)$$

3. Lift: This criterion, also known as "Intersect Factor" or "Interestingness," shows the level of independence between objects A and B. The value of the criterion can be a numerical value between 0 to infinity. The values less than 1 indicate the negative relationship between A and B; hence, the extracted law is not attractive. The values higher than 1 indicate that A provides more information about B, causing a higher attractiveness of the evaluation law. The lift criterion for A and B is shown in Eq. (26) and Eq. (27) as the following:

$$\text{Lift}(A \Rightarrow 0) = \frac{P(A|0)}{P(0)} = \frac{P(A \cap 0)}{P(A) \cdot P(0)} \quad (26)$$

$$\text{Lift}(B \Rightarrow 0) = \frac{P(1|B)}{P(1)} = \frac{P(B \cap 1)}{P(B) \cdot P(1)} \quad (27)$$

For instance, suppose that the behavior of user X is as mentioned below as well as this user has rated the items according to Table 3 and provided the corresponding score for each in the range of [0-5] as follows:

**Table 3. An example of the behavior of user X**

| User X | 1 | 2 | 3 | 4 | 5 | 6 | 7 | 8 | 9 | 10 |
|---|---|---|---|---|---|---|---|---|---|---|
| **Items** | 27 | 33 | 115 | 178 | 203 | 240 | 259 | 307 | 333 | 377 |
| **Ratings** | 3 | 4 | 2 | 4 | 5 | 5 | 4 | 3 | 4 | 3 |

As can be seen, the user has rated ten items and given the corresponding score to each of the items. In the beginning, the average scores given by users are calculated according to Eq. (20) to determine the minimum scores for the interest range of users, which is equal to (Interest Usex X → ⌊3.7⌋ = 3 ). Accordingly, the items with a minimum score of 3 are in the range of the user's interests. In the second step, the association rules are discussed, and the behavior of nodes that were visited in the trusted network and did not rate the considered item is compared with the user X, and the results are presented in Figure 1.

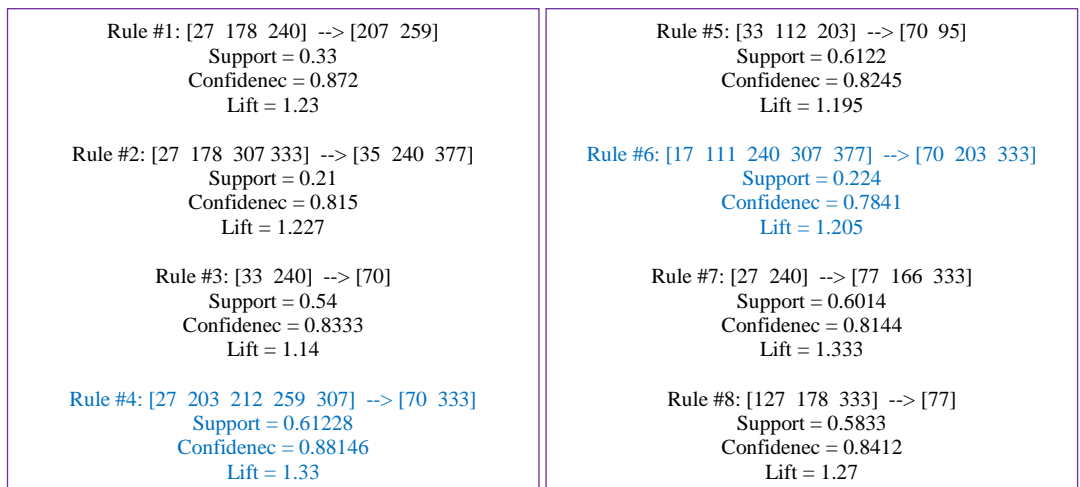

Rule #1: [27 178 240] --> [207 259]
Support = 0.33
Confidenec = 0.872
Lift = 1.23

Rule #2: [27 178 307 333] --> [35 240 377]
Support = 0.21
Confidenec = 0.815
Lift = 1.227

Rule #3: [33 240] --> [70]
Support = 0.54
Confidenec = 0.8333
Lift = 1.14

Rule #4: [27 203 212 259 307] --> [70 333]
Support = 0.61228
Confidenec = 0.88146
Lift = 1.33

Rule #5: [33 112 203] --> [70 95]
Support = 0.6122
Confidenec = 0.8245
Lift = 1.195

Rule #6: [17 111 240 307 377] --> [70 203 333]
Support = 0.224
Confidenec = 0.7841
Lift = 1.205

Rule #7: [27 240] --> [77 166 333]
Support = 0.6014
Confidenec = 0.8144
Lift = 1.333

Rule #8: [127 178 333] --> [77]
Support = 0.5833
Confidenec = 0.8412
Lift = 1.27

**Figure 1. The output obtained from the implementation of the association rules algorithm**

As shown in Figure 1, the relevant rules are extracted, and the most item repetitions are in Rules 4 and 6 (highlighted in blue), which include five repeated items. The users' behavioral similarity with the user X is in 4 and 3 items in rules 4 and 6, respectively. The related items are also shown besides these items, which include the items 70,203,333. Therefore, it can be concluded that item 70 is related to the other five items. Thus, the score given to the item 70 is evaluated, and if its minimum score is equal to 3, it can be recommended to user X. If the score of item 70 is less than 3, other items extracted in the next rules can be discussed, and various recommendations can be made to the target user.

## 4. Proposed Method

The present research objective is to discover and predict the score of items that are considered unknown and were not rated by the target user, and finally, recommend the item to the user or items that have a friendly relationship with the considered item. For this purpose, the search is performed for the users who are trusted by the target user. The desired trust is defined and evaluated by a number of criteria, and then the recommendations related to the target user are presented based on the output. The steps of this operation in the proposed method are as presented in Figure 2.

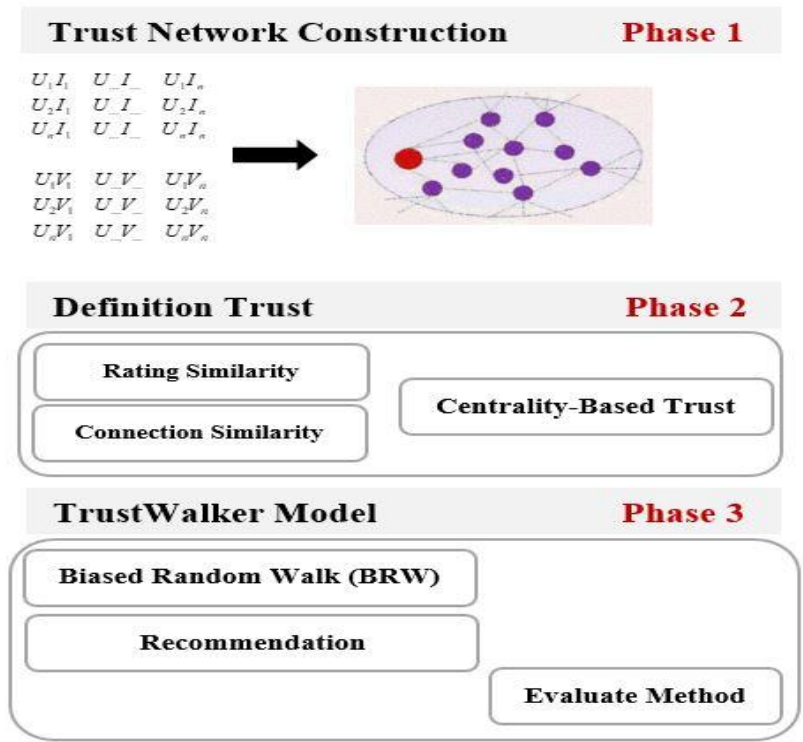

**Figure 2. The framework of the proposed method**

As demonstrated, the proposed method consists of three phases. In the first phase, a trusted network is created for the target user, and the trusted network also consists of two user-item (behavioral similarity based on item selection) and user-user (possessing mutual friends) matrices and user-user. The trust network is created based on these two matrices, and the relationships between the users and the target user as well as other users are determined., In the second phase, after developing a trusted network, the trust is defined. The second phase also includes the calculation of rating similarity, connection similarity, and centrality-based trust. As mentioned earlier, in the present study, it is assumed that the degree of trust between users is obtained through the degree of similarity between them as well as the importance of users in social networks. To evaluate the similarity between users, the similarity between the items rated by them and also the degree of similarity in their connections (the number of friends) is applied, and also the importance of users in the social network is obtained based on the definition of a centrality which includes the definition of H-index and its improvement. Accordingly, the degree of trust between users in the trusted network created for the user is obtained by assessing the mentioned factors. The ways of evaluating and calculating each of these criteria for creating a trust has been fully explained in Section 3. In the third phase, the TrustWalker Model is also presented, and according to the results obtained from the first and second phases, it starts from the target user to visit the nodes using the Biased Random Walk (BRW) algorithm. In this phase, the movements between nodes will be affected by the weight of edges (based on three criteria defined for trust). In the following, some recommendations are presented to the target user according to the visited nodes and the obtained results. If the score of the searched item is available among the visited users, that item will be returned to the target user, and if the visited user has not rated that item, the behavioral similarity between the visited users and the target user is evaluated using association rules, and if there is a behavioral similarity, the items dependent to the rated items by the target user are recommended to the user, which is fully discussed in Section 3.5. Finally, the proposed method is evaluated, which includes the comparison of this method with other approaches. The algorithm of the proposed method is presented as Algorithm 1:

**Algorithm 1.**
**Require:** Target User $U_t$

1: $U_C \leftarrow U_t$
2:     **while** the walk is not terminated **do**
3:   Assign a value to edges based on Trusted users:
4:     Calculate $(2 \times \alpha_1) + \alpha_2 + \alpha_3$   // formula 18
5:   **S** = Select one of the Active User (**n**) neighbors by TrustWalker
6:     **If** S has a rating for target item then
7:   Return the rating value
8:   **end**
9: **end**
10:  Calculate $P_{i,u,k}$   // formula 11
11:    R = Random Number
12:     **If** $R < P_{i,u,k}$ && $Sim_{Pea}(a,b) \succ 0$  // Formula 1 && Steps = Max_Step  // 6 degree
13:   Calculate Association Rules
14:     **else if**
15:      $R < P_{i,u,k}$ && $Sim_{Pea}(a,b) \prec 0$ && Steps > Max_Step
16:   Return 0 or 1   // formula 15 and 16
17:     **while** Steps < Max_Step
18:  Calculate step 10
19:  **else**
20:    Return "**Can not cover**"
21:     **end**
22:     **end**
23:  **end**

# 5. Experiments
## 5.1. Datasets
In the present research, in order to evaluate the proposed method, the datasets of the Ref. [57] were used, and the descriptions provided for each of the datasets are as follows:

In this paper, three real-world datasets including Epinions[1], Flixster[2], FilmTrust[3] are used in the experiments [57]. Epinions is a website in which users can express their opinions about items (such as movies, books, and software) by assigning numerical ratings and writing text reviews. Users can specify other users as trustworthy (to the trust list) or untrustworthy (to the distrust list) according to whether the text reviews and comments of other users are consistently valuable to them or not. The data set is generated by [36], consisting of 49 K users who issued 664 K ratings over 139 K different items and 478 K trust statements. The ratings are integers ranged from 1 to 5, and the trust values are also binary (either 1 or 0).

Flixster is a social movie site in which users are allowed to share their movie ratings, discover new movies, and interact with others who have similar tastes. We adopt the data set[4] collected by [15] which includes a large amount of data. The ratings are real values ranged from 0.5 to 4.0 with an interval 0.5, and the trust statements are scaled from 1 to 10 but not available. Hence, they are converted into binary values the same as FilmTrust, that is, trust value 1 is assigned to a user who is identified as a trusted neighbor and 0 otherwise. Note that the trust statements in this data set is symmetric. We sample a subset by randomly choosing 53 K users who issued 410K item ratings and 655 K trust ratings.

---

[1] http://www.epinions.com

[2] http://www.flixster.com

[3] http://trust.mindswap.org/FilmTrust

[4] http://www.cs.sfu.ca/sja25/personal/datasets

FilmTrust is a trust-based social site in which users can rate and review movies. Since there are no publicly available data sets due to the preservation of user privacy, we crawled the whole site in June 2011, collecting 1986 users, 2071 movies, and 35,497 ratings. The ratings take values from 0.5 to 4.0 with step 0.5. In addition, we also gathered 1853 trust ratings that are issued by 609 users. The average number of trusted neighbors per user is less than 1. Originally, users can specify other users as trusted neighbors with a certain level of trust from 1 to 10. However, these trust values are not available due to the sharing policy. We can only get the link information among users and hence the trust value is 1 if a link exists between two users otherwise the value is 0.

It is noted that all the data sets are highly sparse, i.e., users only rate a small portion of items in the system. The rating sparsity is computed by:

$$\text{Sparsity} = 1 - (\frac{Ratings}{Users \times Items}) \times 100 \quad (28)$$

The information about each of the datasets is presented in Table 3.

**Table 3. The information about the studied datasets**

| Data set | Users | Items | Ratings | Trust | Sparsity (%) |
|---|---|---|---|---|---|
| *Epinions* | 49K | 139K | 664K | 478K | 99.95 |
| *Flixster* | 53K | 18K | 410K | 650K | 99.96 |
| *FilmTrust* | 1986 | 2071 | 35,497 | 1853 | 99.86 |

## 5.2. Evaluation Measures

Leave-One-Out (LOO) method has been extensively used to evaluate recommender systems [15, 58, 59]. In the present article, LOO is applied to assess the proposed method, and the precision of different algorithms in predicting the score of the items is compared by hiding the actual scores. Mean Absolute Error (MAE) and Root Mean Squared Error (RMSE) have been used in various studies [15, 20, 59]. The calculation of error of the recommender systems using MAE and RMSE is performed by the following equations:

$$MAE = \frac{\sum_{i=1}^{n} |R_{u,s} - R'_{u,s}|}{N} \quad (29)$$

$$RMSE = \sqrt{\frac{\sum_{u,s}(R_{u,s} - R'_{u,s})}{N}} \quad (30)$$

Where, $R_{u,s}$ implies the actual score of the items rated by the user and $R'_{u,s}$ is the predicted score for the items by the recommender system. N is the total number of predictions made by the recommender system. The lower the value obtained for both equations, the lower the algorithms' error in predicting the score of items; hence, the algorithm with the lowest error is more precise in predicting the score of items. Since the evaluated data are limited, the recommender system may not be capable of predicting the score of items for all users and items due to data scattering and lack of information in some cases. In order to predict the score of items and calculate the similarity between users, the relevant information of each user should be available that includes <*User: Centrality, Item*>, and if the recommender system does not have this information, it will be unable to cover this user. Hence:

$$\text{Coverage} = \frac{S}{N} \times 100 \quad (31)$$

Where, S implies the number of predicted scores and N is the number of tested scores, and it indicates how many scores the recommender system could cover and test the data for each Coverage dataset.

Precision: In this context, precision refers to a normalized form of RMSA and is defined as follows:

$$\text{Precision} = 1 - \frac{RMSE}{RMSE_{Max}} \quad (32)$$

In this regard, for the Epinions dataset, the score range is between [1,5], and the value of $RMSE_{Max} = 4$ is considered as the maximum possible error [15, 59]. For the Flixster dataset, the score range is between [0.5,4.0], and the value of $RMSE_{Max} = 3$ is considered as the maximum possible error. In this context, for the FilmTrust dataset, the score range is between [1,10], and the value of $RMSE_{Max} = 9$ is considered as the maximum possible error. F-Measure is the next criterion to be evaluated and is defined as follows:

$$\text{F-Measure} = \frac{2 \times \text{Precision} \times \text{Coverage}}{\text{Precision} + \text{Coverage}} \quad (33)$$

### 5.3. Comparing with Other Methods

In this section, the proposed method is compared with other approaches based on the considered criteria, and the efficiency of each method is evaluated. The compared methods include CFPearson [60], ItemCF [61], TidalTrust [58], MoleTrust [36], TrustWalker [15], CoTrustWalker [62], CliquesWalker [63], TrustMF [54], TrustSVD [40], and Trust-Enhanced [59] and the results have been obtained based on the criteria considered from the comparison of the proposed algorithm called Centrality Connection Items-TrustWalker (CCI-TrustWalker) with other algorithms and on the stated datasets. The users have also been classified into four categories: 25%, 50%, 75%, and 100%. The MAE and RMSE criteria are implemented for each of the mentioned intervals, and the results are presented in Tables 4, 5, and 6 as follows:

**Table 4. Epinions dataset MAE and RMSE**

| | EPINIONS | 25% | | 50% | | 75% | | 100% | |
|---|---|---|---|---|---|---|---|---|---|
| | | MAE | RMSE | MAE | RMSE | MAE | RMSE | MAE | RMSE |
| 1 | CFPearson | 0.8752 | 0.9355 | 0.8326 | 0.9124 | 0.8157 | 0.9031 | 0.8517 | 0.9228 |
| 2 | ItemCF | 0.8122 | 0.9011 | 0.8375 | 0.9151 | 0.8144 | 0.9024 | 0.7672 | 0.8758 |
| 3 | TidalTrust | 0.7711 | 0.8781 | 0.7955 | 0.8919 | 0.8149 | 0.9027 | 0.7147 | 0.8453 |
| 4 | MoleTrust | 0.8221 | 0.9066 | 0.7433 | 0.8621 | 0.7315 | 0.8552 | 0.8176 | 0.9042 |
| 5 | TrustWalker | 0.7948 | 0.8915 | 0.7339 | 0.8566 | 0.6552 | 0.8094 | 0.6773 | 0.8229 |
| 6 | CoTrustWalker | 0.6525 | 0.8077 | 0.6725 | 0.8200 | 0.5335 | 0.7304 | 0.5426 | 0.7366 |
| 7 | CliquesWalker | 0.6438 | 0.8023 | 0.6339 | 0.7961 | 0.6019 | 0.7758 | 0.5202 | 0.7212 |
| 8 | TrustMF | 0.6178 | 0.7860 | 0.7224 | 0.8499 | 0.6064 | 0.7787 | 0.5925 | 0.7697 |
| 9 | TrustSVD | 0.6517 | 0.8072 | 0.6104 | 0.7812 | 0.5823 | 0.7630 | 0.5546 | 0.7447 |
| 10 | Trust-Enhanced | 0.5209 | 0.7217 | 0.5147 | 0.7174 | 0.4887 | 0.6990 | 0.5052 | 0.7107 |
| 11 | CCI- TrustWalker | 0.4141 | 0.6435 | 0.3702 | 0.6084 | 0.3555 | 0.5962 | 0.3324 | 0.5765 |

**Table 5. Flixster dataset MAE and RMSE**

| | FLIXSTER | 25% | | 50% | | 75% | | 100% | |
|---|---|---|---|---|---|---|---|---|---|
| | | MAE | RMSE | MAE | RMSE | MAE | RMSE | MAE | RMSE |
| 1 | CFPearson | 0.8996 | 0.9484 | 0.8502 | 0.9220 | 0.8144 | 0.9024 | 0.7224 | 0.8499 |
| 2 | ItemCF | 0.8577 | 0.9261 | 0.8621 | 0.9284 | 0.8225 | 0.9069 | 0.7416 | 0.8611 |
| 3 | TidalTrust | 0.8100 | 0.9000 | 0.7855 | 0.8862 | 0.7362 | 0.8580 | 0.7446 | 0.8629 |
| 4 | MoleTrust | 0.7663 | 0.8753 | 0.7230 | 0.8502 | 0.7495 | 0.8657 | 0.6844 | 0.8272 |
| 5 | TrustWalker | 0.7544 | 0.8685 | 0.7108 | 0.8430 | 0.7233 | 0.8504 | 0.6430 | 0.8018 |
| 6 | CoTrustWalker | 0.8004 | 0.8946 | 0.8125 | 0.9013 | 0.7150 | 0.8455 | 0.6314 | 0.7946 |
| 7 | CliquesWalker | 0.7436 | 0.8623 | 0.8004 | 0.8946 | 0.7364 | 0.8581 | 0.6666 | 0.8164 |
| 8 | TrustMF | 0.6547 | 0.8091 | 0.6320 | 0.7949 | 0.6500 | 0.8062 | 0.5901 | 0.7681 |
| 9 | TrustSVD | 0.5669 | 0.7529 | 0.5435 | 0.7372 | 0.5514 | 0.7425 | 0.5107 | 0.7146 |
| 10 | Trust-Enhanced | 0.5503 | 0.7418 | 0.5127 | 0.7160 | 0.5001 | 0.7071 | 0.4667 | 0.6831 |
| 11 | CCI- TrustWalker | 0.4217 | 0.6493 | 0.4012 | 0.6334 | 0.3879 | 0.6228 | 0.3547 | 0.5955 |

**Table 6. FilmTrust dataset MAE and RMSE**

| | FILMTRUST | 25% | | 50% | | 75% | | 100% | |
|---|---|---|---|---|---|---|---|---|---|
| | | MAE | RMSE | MAE | RMSE | MAE | RMSE | MAE | RMSE |
| 1 | CFPearson | 0.8604 | 0.9275 | 0.8210 | 0.9060 | 0.7714 | 0.8782 | 0.6743 | 0.8211 |
| 2 | ItemCF | 0.8021 | 0.8956 | 0.7422 | 0.8615 | 0.8055 | 0.8974 | 0.7336 | 0.8565 |
| 3 | TidalTrust | 0.7816 | 0.8840 | 0.7314 | 0.8552 | 0.6825 | 0.8261 | 0.7100 | 0.8426 |
| 4 | MoleTrust | 0.8210 | 0.9060 | 0.8354 | 0.9140 | 0.7438 | 0.8624 | 0.6775 | 0.8231 |
| 5 | TrustWalker | 0.7652 | 0.8747 | 0.7750 | 0.8803 | 0.7229 | 0.8502 | 0.6744 | 0.8212 |
| 6 | CoTrustWalker | 0.7324 | 0.8558 | 0.6655 | 0.8157 | 0.6111 | 0.7817 | 0.6332 | 0.7957 |
| 7 | CliquesWalker | 0.6419 | 0.8011 | 0.5918 | 0.7692 | 0.5554 | 0.7452 | 0.5104 | 0.7144 |
| 8 | TrustMF | 0.7116 | 0.8435 | 0.6413 | 0.8008 | 0.5633 | 0.7505 | 0.5764 | 0.7592 |
| 9 | TrustSVD | 0.6300 | 0.7937 | 0.5863 | 0.7657 | 0.4818 | 0.6941 | 0.4580 | 0.6767 |
| 10 | Trust-Enhanced | 0.5867 | 0.7659 | 0.5240 | 0.7238 | 0.4633 | 0.6806 | 0.4166 | 0.6454 |
| 11 | CCI- TrustWalker | 0.4829 | 0.6949 | 0.4355 | 0.6599 | 0.3691 | 0.6075 | 0.3417 | 0.5845 |

As can be seen, the CCI-TrustWalker algorithm has the lowest error in all three datasets compared to other algorithms. In the following, the precision criterion is calculated, and the accuracy of the algorithms is calculated according to the RMSE value obtained for each of the algorithms. Also, the results obtained for each algorithm are compared and presented in Table 7 as follows:

**Table 7. Epinions, Flixster, and FilmTrust datasets Precision**

| | | Precision | 25% | 50% | 75% | 100% |
|---|---|---|---|---|---|---|
| Epinions | | | | | | |
| Flixster | | | | | | |
| FilmTrust | | | | | | |
| | | | 0.7661 | 0.7719 | 0.7742 | **0.7693** |
| | 1 | CFPearson | 0.6838 | 0.6926 | 0.6992 | **0.7183** |
| | | | 0.8969 | 0.8993 | 0.9024 | **0.9087** |
| | | | 0.7747 | 0.7712 | 0.7744 | **0.7810** |
| | 2 | ItemCF | 0.6913 | 0.6905 | 0.6977 | **0.7129** |
| | | | 0.9004 | 0.9042 | 0.9002 | **0.9048** |
| | | | 0.7804 | 0.7770 | 0.7743 | **0.7886** |
| | 3 | TidalTrust | 0.7000 | 0.7046 | **0.7140** | 0.7123 |
| | | | 0.9017 | 0.9049 | **0.9082** | 0.9063 |
| | | | 0.7733 | 0.7844 | **0.7862** | 0.7744 |
| | 4 | MoleTrust | 0.7082 | 0.7166 | 0.7114 | **0.7242** |
| | | | 0.8993 | 0.8984 | 0.9041 | **0.9085** |
| | | | 0.7771 | 0.7858 | **0.7976** | 0.7942 |
| | 5 | TrustWalker | 0.7105 | 0.7190 | 0.7165 | **0.7327** |
| | | | 0.9028 | 0.9021 | 0.9055 | **0.9087** |
| | | | 0.7980 | 0.7950 | **0.8174** | 0.8158 |
| | 6 | CoTrustWalker | 0.7018 | 0.6995 | 0.7181 | **0.7351** |
| | | | 0.9049 | 0.9093 | **0.9131** | 0.9115 |
| | | | 0.7994 | 0.8009 | 0.8060 | **0.8197** |
| | 7 | CliquesWalker | 0.7125 | 0.7018 | 0.7139 | **0.7278** |
| | | | 0.9109 | 0.9145 | 0.9172 | **0.9206** |

| # | Algorithm | | | | |
|---|---|---|---|---|---|
| 8 | TrustMF | 0.8035 | 0.7875 | 0.8053 | **0.8075** |
| | | 0.7303 | 0.7350 | 0.7312 | **0.7439** |
| | | 0.9062 | 0.9110 | **0.9166** | 0.9156 |
| 9 | TrustSVD | 0.7982 | 0.8047 | 0.8092 | **0.8138** |
| | | 0.7490 | 0.7542 | 0.7525 | **0.7618** |
| | | 0.9118 | 0.9149 | 0.9228 | **0.9248** |
| 10 | Trust-Enhanced | 0.8195 | 0.8206 | **0.8252** | 0.8223 |
| | | 0.7527 | 0.7613 | 0.7643 | **0.7723** |
| | | 0.9149 | 0.9195 | 0.9243 | **0.9282** |
| 11 | CCI- TrustWalker | 0.8391 | 0.8479 | 0.8509 | **0.8558** |
| | | 0.7835 | 0.7888 | 0.7924 | **0.8015** |
| | | 0.9227 | 0.9266 | 0.9325 | **0.9350** |

In each row of Table 3, each of the algorithms has been implemented for different intervals in all three datasets, and each dataset is marked with a separate color. Also, in each row, Precision's maximum value is highlighted in yellow for each implementation of the relevant algorithm in each dataset. As can be seen, the maximum value of Precision in all three datasets is related to the CCI-TrustWalker algorithm. Finally, the algorithms are evaluated using the F-Measure criterion. Before performing this action, it is needed to calculate the Coverage for each of the algorithms in all three datasets, the results of which are indicated in Figure 3 as follows:

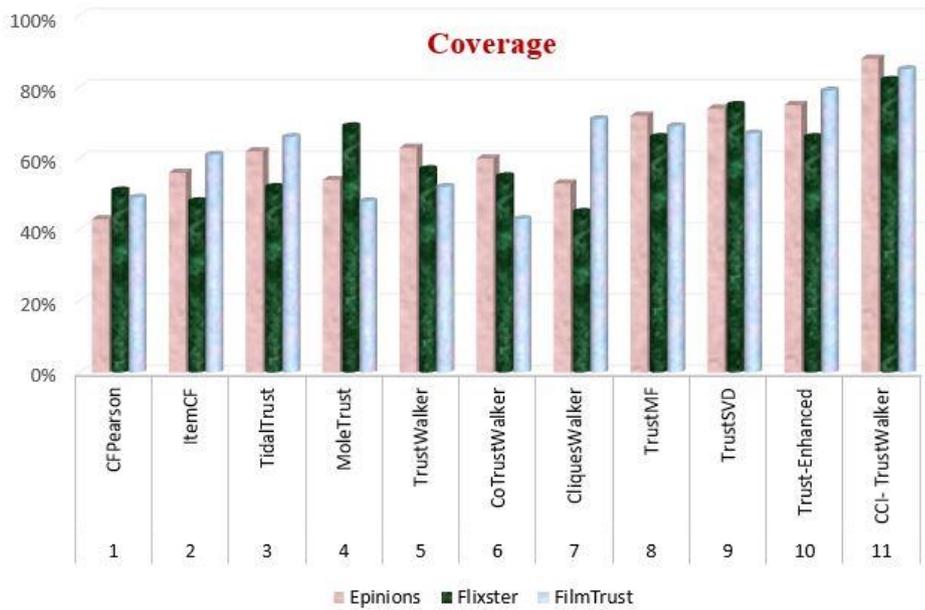

**Fig 3. Algorithms Coverage in the datasets**

As evidenced, the CCI-TrustWalker algorithm had the highest Coverage in all three datasets compared to other algorithms. The F-Measure criterion is calculated based on the obtained Coverage and Precision, and the results are presented in Table 8.

**Table 8. Epinions, Flixster, and FilmTrust datasets F–Measure**

| | | F–Measure | 25% | 50% | 75% | 100% |
|---|---|---|---|---|---|---|
| | | | Epinions | | | |
| | | | Flixster | | | |
| | | | FilmTrust | | | |
| 1 | CFPearson | | 0.5508 | 0.5523 | **0.5529** | 0.5516 |
| | | | 0.5842 | 0.5874 | 0.5897 | **0.5964** |
| | | | 0.6337 | 0.6343 | 0.6351 | **0.6366** |
| 2 | ItemCF | | 0.6500 | 0.6488 | 0.6499 | **0.6522** |
| | | | 0.5665 | 0.5663 | 0.5682 | **0.5737** |
| | | | 0.7272 | 0.7285 | 0.7272 | **0.7287** |
| 3 | TidalTrust | | 0.6910 | 0.6896 | 0.6886 | **0.6942** |
| | | | 0.5967 | 0.5983 | **0.6017** | 0.6011 |
| | | | 0.7621 | 0.7632 | **0.7644** | 0.7637 |
| 4 | MoleTrust | | 0.6359 | 0.6396 | **0.6402** | 0.6360 |

| | | | | | |
|---|---|---|---|---|---|
| | | 0.6989 | 0.7030 | 0.7005 | **0.7066** |
| | | 0.6259 | 0.6256 | 0.6270 | **0.6281** |
| | | 0.6958 | 0.6993 | **0.7039** | 0.7026 |
| 5 | TrustWalker | 0.6325 | 0.6401 | 0.6349 | **0.6411** |
| | | 0.6599 | 0.6597 | 0.6606 | **0.6614** |
| | | 0.6849 | 0.6838 | **0.6920** | 0.6914 |
| 6 | CoTrustWalker | 0.6166 | 0.6158 | 0.6229 | **0.6292** |
| | | 0.5829 | 0.5838 | **0.5846** | 0.5843 |
| | | 0.6374 | 0.6378 | 0.6394 | **0.6437** |
| 7 | CliquesWalker | 0.5516 | 0.5483 | 0.5520 | **0.5561** |
| | | 0.7979 | 0.7993 | 0.8004 | **0.8016** |
| | | 0.7594 | 0.7522 | 0.7602 | **0.7612** |
| 8 | TrustMF | 0.6933 | 0.6954 | 0.6937 | **0.6994** |
| | | 0.7834 | 0.7852 | **0.7873** | 0.7869 |
| | | 0.7679 | 0.7709 | 0.7730 | **0.7751** |
| 9 | TrustSVD | 0.7497 | 0.7520 | 0.7512 | **0.7558** |
| | | 0.7724 | 0.7735 | 0.7763 | **0.7770** |
| | | 0.7832 | 0.7837 | **0.7858** | 0.7844 |
| 10 | Trust-Enhanced | 0.7033 | 0.7070 | 0.7083 | **0.7174** |
| | | 0.8478 | 0.8498 | 0.8518 | **0.8535** |
| | | 0.8590 | 0.8636 | 0.8652 | **0.8677** |
| 11 | CCI- TrustWalker | 0.8013 | 0.8040 | 0.8059 | **0.8106** |
| | | 0.8848 | 0.8866 | 0.8893 | **0.8904** |

As demonstrated, the maximum value for each of the algorithms and in each dataset is highlighted in yellow color. The results presented in Table 4 in all three datasets indicate that the method presented in the present study has higher efficiency than other algorithms based on the F-Measure criterion.

## 6. Conclusions

In the present article, the scores of items that the user has not rated were predicted, and a trust-based recommender system was used to predict the score of these items. For this purpose, a trusted network is created that includes users who have similarities in behaviors with the target user in the selection of items and friends. After creating the trusted network, a TrustWalker is developed that randomly selects the nodes in the trusted network by employing the BRW algorithm. Before the movement of TrustWalker between users on the network, the degree of trust between them is defined and calculated. In order to define trust between users, three criteria are applied that include the similarity of the selection of items by users, the similarity in the connections (mutual friends), and also the importance of a node in the social network. Centrality has been defined for each node, and H-index has been used and improved to evaluate the node's importance in the social network.

After calculating each of these criteria, the results have been used to weigh the nodes' edges. The greater the values for weights of edges, the higher the degree of trust between the two nodes with which the edge is connected. After weighting the edges, the BRW algorithm is implemented, which starts moving from the source node and visits the node at each random step. The weight of each edge affects the selection of the BRW random step. TrustWalker evaluates two modes for terminating the walk on the trusted network. An important point in the implementation of the proposed method is that in the weighting of the edges between the nodes, the degree of behavioral similarity between the nodes is considered doubled compared to the other two criteria with impact factors, and this leads to more probability of the selection of users with behavioral similarity with the target user. It also makes it possible to discover the items dependent on the target user's interests using the association rules and behavioral similarity between them in the case that the searched item is not rated by visited users.

The proposed recommender system also acts in two modes. In the first mode, if the system finds the score of the desired item, returns it to the target user, and in the second mode, if it does not detect the score of the item, it finds some items and recommends that these items have dependencies with the selected items by the target user. The results obtained from the output of the proposed method in the present research, in comparison with other methods in the evaluation section, indicate better performance of this algorithm. The proposed method has also caused the BRW to move toward nodes that are trustful to the source user. This trust itself is obtained from the degree of similarity between users and their importance in the social network, making the recommender system either find the score of the searched item immediately or make some recommendations to the users in the range of their interests.

## Acknowledgments

The authors thank Professor Pavel N. Krivitsky (Department of Statistics, School of Mathematics and Statistics, UNSW Sydney, NSW, Australia) for his invaluable suggestions and skilled technical assistance.